\documentclass[aip,jap,showpacs,a4paper,showkeys,amssymb,amsmath,
superscriptaddress, floatfix,
reprint]
{revtex4-1}

\usepackage{graphicx}
\usepackage{amsmath}
\usepackage{amssymb}
\usepackage{color}
\usepackage{verbatim}
\usepackage{url}
\usepackage{bm}
\usepackage{multirow}
\usepackage{rotating}

\renewcommand{\vec}[1]{\bm{#1}}
\newcommand{\ket}[1]{|{#1} \rangle}
\newcommand{\bra}[1]{\langle {#1}|}

\begin{document}

\title {Multiscale model for phonon-assisted band-to-band tunneling in
semiconductors}

\author{Arvind Ajoy}
\email[]{arvindajoy@iitm.ac.in}
\affiliation{\mbox{Department   of   Electrical  Engineering,   Indian
Institute of Technology Madras, Chennai 600036, India}}

\author{S. E. Laux}
\email[]{laux@us.ibm.com}
\affiliation{IBM T.  J. Watson Research Center,  Yorktown Heights, New
York 10598, USA}

\author{Kota V. R. M. Murali}
\email[]{kotamurali@in.ibm.com}
\affiliation{\mbox{IBM Semiconductor  Research and Development Center,
Bangalore 560045, India}}

\author{Shreepad Karmalkar}
\email[]{karmal@ee.iitm.ac.in}
\affiliation{\mbox{Department of Electrical Engineering, Indian Institute of
Technology Madras, Chennai 600036, India}}

\date{\today}

\begin{abstract}
We  present  a TCAD  compatible  multiscale  model of  phonon-assisted
band-to-band tunneling (BTBT) in semiconductors, that incorporates the
non-parabolic  nature  of complex  bands  within  the  bandgap of  the
material.   This model  is shown  capture  the measured
current-voltage  data  in silicon,  for  current  transport along  the
$[100]$, $[110]$ and  $[111]$ directions. Our model will  be useful to
predict  band-to-band  tunneling  phenomena  to quantify  on  and  off
currents in Tunnel FETs and in small geometry MOSFETs and FINFETs.
\end{abstract}


\maketitle

\section{Introduction}
The on-current in Tunnel FETs and gate induced off-state drain current
in  small geometry  MOSFETs  are  due to  the  tunneling of  electrons
between valence and conduction bands. This work deals with the process
of  phonon-assisted band-to-band tunneling  (BTBT) across  an indirect
bandgap.   One approach  to  compute phonon-assisted  BTBT current  is
based on  the Non-Equilibrium  Green's function (NEGF)  technique, for
e.g.   Refs.  \onlinecite  {Rivas_APL_2001, Luisier_JAP_2010}  using a
basis of atomic orbitals.   Electron transport is not ballistic, since
scattering due to phonons is  the driving force for BTBT current.  The
atomistic NEGF  approach, though  rigorous and accurate,  requires the
use    of   supercomputers    \cite{Luisier_HPC_2008}    to   simulate
realistically   sized   devices,  especially   when   the  effect   of
electron-phonon  coupling \cite{Luisier_PRB_2009}  is  included.  More
efficient quantum transport algorithms such as the Wavefunction Method
\cite{Luisier_PRB_2006}  cannot be used  since scattering  is present.
An  alternate  approach is  to  use  the conventional  drift-diffusion
equations of semiconductor transport  with a suitably calibrated model
(eg.       Refs.      \onlinecite{Hurkx_TED_1992,     Pandey_TED_2010,
Kao_IEEE_TED_2012})  describing   the  process  of   tunneling.   Most
commercially  available  semiconductor  device simulators  (TCAD)  are
based on this latter approach.

BTBT occurs via evanescent  states corresponding to the conduction and
valence bands.   The properties of evanescent states  are described by
the complex bandstructure of  the material. TCAD compatible models for
BTBT  in  an  indirect  bandgap  semiconductor  \cite{Kane_JAP_1961, Tanaka_SSE_1994,
Vandenberghe_JAP_2011,   Schenk_SSE_1993,  Keldysh_JETP_1958}   use  a
simple  parabolic approximation for  the complex  bandstructure within
the bandgap,  since the curvatures of  the real and  complex bands are
identical    at     the    band    extrema    \cite{Kohn_PhysRev_1963,
Heine_ProcPhysSoc_1963}.   However, this  approximation  can introduce
large  errors in  BTBT currents,  since the  tunneling  current depends
exponentially on  the action for  tunneling, which in turn  depends on
the complex bandstructure  over the entire bandgap, not  merely at the
band  extrema  (see   Ref.  \onlinecite{Laux_IWCE_2009}  and  Section
\ref{Sec_Parabolic}).   A  first  attempt  to include  the  effect  of
non-parabolic complex bands to compute BTBT across an indirect bandgap
\cite{Pandey_TED_2010}  ignored  the role  of  phonons,  and used  the
Esaki-Tsu  formula \cite{Tsu_APL_1973}  meant  for electron  tunneling
between conduction bands, leading  to a prefactor which is independent
of  the  valence  band   effective  mass.   We  present  a  physically
consistent,    multiscale   model    that   incorporates    both   the
non-parabolicity  of  the  complex   bands  and  the  physics  of  the
electron-phonon interaction.   The non-parabolicity is  captured using
energy   dependent  effective  masses   \cite{Pandey_TED_2010},  which
connect  a computation  carried out  on an  atomistic scale  (using an
$sp^3d^5s^*$  tight binding  scheme) with  a tunneling  model  that is
formulated using effective mass  wave functions describing much larger
length scales.  Our model is symmetric with respect to the valence and
conduction band parameters.  This model can easily be implemented in a
conventional  TCAD tool.  Finally,  our model  is shown  to 
capture the  measured current-voltage data  \cite{Solomon_DRC_2009} in
silicon for  current transport along the $[100]$,  $[110]$ and $[111]$
directions.

This paper  is organized as  follows.  In section  \ref{Sec_Model}, we
describe   and   derive    the   multiscale   BTBT   model.    Section
\ref{Sec_Results} compares the results  of our model with experimental
data. Section \ref{Sec_Parabolic} demonstrates the inadequacy of using
a parabolic  approximation to the  complex bands while  computing BTBT
currents.   Section   \ref{Sec_Conclusion}  summarizes  the  important
conclusions.    Finally,    the   appendices   provide   supplementary
information that will be useful to implement our model.

\section{Model}
\label{Sec_Model}
Our   approach    is   motivated    by   a   combination    of   Refs.
\onlinecite{Tanaka_SSE_1994, Pandey_TED_2010,  Laux_xxx}.  We restrict
our attention  to a 1-D  problem. For definiteness, let  $x$ represent
the   transport  direction.    Then,  in   brief,   Ref.   \onlinecite
{Tanaka_SSE_1994} uses a simple WKB form ($\sim e^{\iota S(x)/\hbar}$,
where the  action $S(x)$ is  correct up to  $\mathcal{O}(\hbar^0)$) to
describe the electronic wavefunctions within the bandgap, whereas Ref.
\onlinecite{Laux_xxx} improves this description by including the first
order term with respect to $\hbar$  in $S(x)$.  We include the idea of
a     position     dependent      effective     mass     from     Ref.
\onlinecite{Pandey_TED_2010} in Ref.  \onlinecite{Laux_xxx}, and use a
WKB  form ($\sim \big(\partial  E /  \partial k  \big)^{-1/2} e^{\iota
S(x)/\hbar}$, with $\partial E / \partial k$ understood to be position
dependent) appropriate  to this situation  \cite{Geller_PRL_1993}.  It
is useful  to note that this modified  WKB form can be  derived from a
transfer   matrix  method   \cite{Huang_ChinJPhys_2008}   by  ignoring
reflections.   Finally, based  on this  insight, we  use  the transfer
matrix  method  to  correct  for   errors  caused  by  the  WKB  based
approach. Note  that we  do not consider  the non-parabolic  nature of
real energy  bands in  this work. This  allows a simple  evaluation of
integrals  corresponding   to  the  density  of   states  involved  in
tunneling. We  believe that this  is a reasonable  approximation while
computing BTBT currents,  since the density of states  scales as $\sim
\text{mass}^{1.5}$,  unlike the  tunneling  probability which  depends
exponentially on  the effective masses  of the complex bands,  via the
action for tunneling. 

We begin by extracting  energy dependent effective masses $m_{VB}(E)$,
$m_{CB}(E)$ of the imaginary parts of the valence and conduction bands
from  a   computation  \cite{Ajoy_DRC_2011,  Ajoy_JPCM_2012}   of  the
direction-dependent     complex     bandstructure    $k^{\parallel}(E;
\vec{k}^{\perp})$ in an $sp^3d^5s^*$ tight binding scheme. The valence
band maxima are  assumed to be at $\vec{k} =  \vec{0}$ to simplify the
description that follows.  Note that $\vec{k}^{\parallel}$ is parallel
or antiparallel  to the transport direction,  and $\vec{k^{\perp}}$ is
chosen by projecting the positions  of all the conduction band valleys
onto the $\vec{k}^{\parallel} = \vec{0}$ plane; $k^{\parallel}$ is the
magnitude of $\vec{k}^{\parallel}$. Note also that we have flipped the
definitions   of   $\perp$   and   $\parallel$  as   used   in   Refs.
\onlinecite{Ajoy_DRC_2011,   Ajoy_JPCM_2012},  in   order   to  remain
consistent with  Ref. \onlinecite{Tanaka_SSE_1994}.  For  each valence
band,  there are  as  many  tunneling paths  as  there are  conduction
valleys, each tagged by a different value of $\vec{k}^{\perp}$. Within
the   bandgap   $E_{VB\;max}   <   E   <   E_{CB\;min}$,   we extract the
masses using the  definitions
$\operatorname{Im}  [k^{\parallel}(E;  \vec{0})]  = \sqrt{2  m_{VB}(E;
\vec{0}  )(E  -  E_{VB  \;  max})}  /  \hbar$  and  $\operatorname{Im}
[k^{\parallel}(E;    \vec{k^{\perp}})    ]    =   \sqrt{2    m_{CB}(E;
\vec{k^{\perp}})(E_{CB  \;  min} -  E)}  /  \hbar$  for the  imaginary
valence and  complex conduction  bands constituting a  tunneling path.
Near the  band edges, the masses  are extracted from  the curvature of
the bands. 

\begin{figure}[!t]
 \centering
   \includegraphics[scale=1.0]{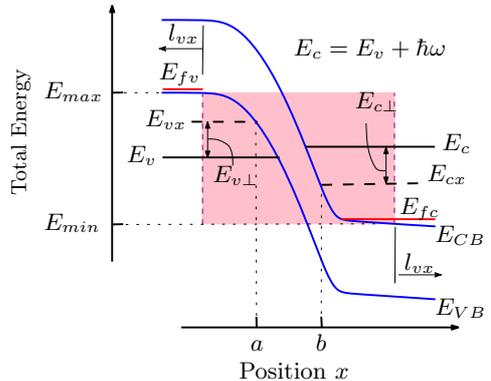}	
\caption{Definition  of  quantities for  BTBT  model, following  Refs.
\onlinecite{Tanaka_SSE_1994, Laux_xxx}.  The  energies $E_c$, $E_v$ as
drawn demonstrate phonon absorption due to electron transfer from $v
\rightarrow  c$. The  shaded region  shows the  tunneling  window. The
potential energy $U(x) \equiv E_{VB}(x)$.}
\label{fig_Tanaka}
\end{figure}

Fig. \ref{fig_Tanaka}  shows the energy band diagram of  a $p\text{-}n$ diode
for a  general case of  non-uniform (and possibly  degenerate) doping.
We consider a large enough  tunneling window so that tunneling current
computed   is    independent   of   its    extent.    Following   Ref.
\onlinecite{Tanaka_SSE_1994}    (also    see    Table   1,    Appendix
\ref{App_Tanaka}),   the  electronic   wavefunction   is  written   as
$\psi(x,y,z) = u_x(x)u_y(y)u_z(z)$, where $u_y(y)$, $u_z(z)$ are plane
waves   with  position-independent   effective  masses   $m_y$,  $m_z$.
The extent of the device in $y$, $z$ directions is denoted
by $l_y$, $l_z$.  An  additional  subscript  $v,c$  is  used  to  denote
quantities on the $p$, $n$ sides of the junction respectively.  Beyond
the classical  turning points  ($x <  a$, $x >  b$), $u_x(x)$  is also
assumed to be a plane wave.  However, we modify the $x$ dependent part
of  the  wavefunctions ($u_{vx}$,  $u_{cx}$  for  an  electron in  the
valence and conduction bands respectively) within the region $ a < x <
b $ to  include the effect of a position  dependent effective mass and
write
\begin{subequations}
\begin{align}
\label{eq_modified_VB}
\nonumber 
 u_{vx}(x) &= \sqrt{ \frac{k_{vx} - k_{0vx}}{|\kappa_{vx}(x)|}
                      \frac{m_{vx}(x)}{m_{vx}} }
               \frac{\exp (\iota k_{0vx}(x-a))}{\sqrt{l_{vx}}} \\
                & \phantom{=} \times
               \exp-\left(\int_a^x \kappa_{vx}(x') dx'\right) \\
\label{eq_modified_CB}
\nonumber  
u_{cx}(x) &= \sqrt{ \frac{k_{cx} - k_{0cx}}{|\kappa_{cx}(x)|}
                      \frac{m_{cx}(x)}{m_{cx}} } 
               \frac{\exp (\iota k_{0cx}(x-b))}{\sqrt{l_{cx}}} \\
               & \phantom{=} \times 
               \exp-\left(\int_x^b \kappa_{cx}(x') dx'\right) 
\end{align}               
\end{subequations}
where $\kappa_{vx}(x)  = \sqrt{2  m_{vx}(x)(E_{vx} - U(x))}  / \hbar$,
and  $\kappa_{cx}(x)  = \sqrt{2  m_{cx}(x)(E_g  +  U(x)  - E_{cx})}  /
\hbar$. Here, $k_{0vx}$, $k_{0cx}$ refer  to the positions of the band
extrema; $l_{vx}$, $l_{cx}$ are the lengths of the regions outside the
tunneling  window   on  the  $p$  and  $n$   sides  respectively  (see
Fig. \ref{fig_Tanaka}); $m_{vx}$, $m_{cx}$ are the effective masses at
the  band edges; and  $m_{vx}(x)$, $m_{cx}(x)$  refer to  the position
dependent   effective  masses  within   the  bandgap,   obtained  from
$m_{VB}(E)$, $m_{CB}(E)$  respectively. The terms  $k_{vx} - k_{0vx}$,
$k_{cx} - k_{0cx}$  are understood to be evaluated at  $x = a^{-}$ and
$x = b^{+}$ respectively.  Note that the products $m_{cx}m_{cy}m_{cz}$
and $m_{vx}m_{vy}m_{vz}$  remain invariant of  the transport direction
\cite{Rahman_JAP_2005}.   We now  follow  the procedure  used in  Ref.
\onlinecite{Tanaka_SSE_1994}.  The essential differences are presented
below.     A   detailed   derivation    is   provided    in   Appendix
\ref{App_Derivation}.

The combined wavefunction of  the electron-phonon system is written as
$\ket{i}  =   \ket{\psi_{i}}  \cdot  \ket{\ldots   n_{\vec{q},  \mu}^i
\dots}$,  where  $i  =  c, v$  and  $  n_{\vec{q},  \mu}^i$  gives  the
occupation   number  of   the  phonon   mode  $\mu$   with  wavevector
$\vec{q}$.     The     electron-phonon     interaction     Hamiltonian
\cite{Tanaka_SSE_1994} is
\begin{align}
\mathcal{W}_{e-ph} = \sum_{\vec{q}, \mu} \frac{M_{\vec{q}, \mu}}{\sqrt{\Omega}}
\left( a_{\vec{q}, \mu} e^{\iota \vec{q} \cdot \vec{r}} + 
a_{\vec{q}, \mu}^{\dagger} e^{-\iota \vec{q} \cdot \vec{r}} \right),
\end{align}
where  $a_{\vec{q},  \mu}^{\dagger},   a_{\vec{q},  \mu}$  are  phonon
destruction, creation operators  and $M_{\vec{q},\mu}$ is the strength
of      the      electron-phonon      interaction.      From      Ref.
\onlinecite{Tanaka_SSE_1995}, $M_{\vec{q}, \mu} = \sqrt{\frac{\hbar}{2
\rho_s  \omega_{\vec{q}}}} D_{\vec{q},  \mu}$, where  $\rho_s$  is the
density  of   the  semiconductor,   and  $D_{\vec{q},  \mu}$   is  the
intervalley deformation potential.  $\hbar \omega_{\vec{q}}$ is energy
of  the phonon  and  $\Omega$ is  the  volume of  the device  $\approx
(l_{cx}+ l_{vx})l_yl_z$.   There are four  processes to be  modeled in
order  to  compute  BTBT  current  -- phonon  emission  or  absorption
(denoted $e/a$)  driving the transfer  of an electron either  from the
valence to the conduction band (denoted $v \rightarrow c$) or from the
conduction   to  the   valence  band   ($c  \rightarrow   v$).   Since
$\mathcal{W}_{e-ph}$  is   Hermitian,  $|  \bra{c}  \mathcal{W}_{e-ph}
\ket{v} |^2  = | \bra{v} \mathcal{W}_{e-ph} \ket{c}  |^2$; i.e.  given
electronic states  with energies  $E_v, E_c$ (that  are assumed  to be
appropriately  filled/empty  to allow  electron  transfer) and  phonon
occupations $n_{\vec{q}, \mu}^{v}, n_{\vec{q}, \mu}^{c}$, the transfer
$\{v \rightarrow  c; e\}$ is equally  likely as $\{c  \rightarrow v; a
\}$ within  the framework  of Fermi's  golden rule. For  want of  a better
alternative, we seek to replace the phonon occupation numbers by their
expectation values  given by the Bose-Einstein  distribution. In doing
so, it is important to recognize  the subtle point that one cannot set
both  $n_{\vec{q},  \mu}^{v}, n_{\vec{q},  \mu}^{c}$  to  be equal  to
$N_{\vec{q}, \mu}$,  the expectation  value. We thus  set $n_{\vec{q},
\mu}^{v}  = N_{\vec{q},  \mu}$ for  a $v  \rightarrow c$  transfer and
$n_{\vec{q},  \mu}^{c} =  N_{\vec{q}, \mu}$  for a  $c  \rightarrow v$
transfer. The implicit assumption is  that there exists a quick phonon
relaxation process  (not modeled by  our Hamiltonian) that  drives the
phonon  population  to  its   equilibrium  value  after  the  electron
transfer.

Consider first the processes $\{v  \rightarrow c; e/a\}$. We then have
the electron phonon interaction as
\begin{subequations}
\label{eq_elec_phon_int1}
\begin{align}
\overline{\bra{c} \mathcal{W}_{e-ph} \ket{v}_{e/a}}
       & = 
      \sum_{q_x, \mu} \left[ R_{e/a} \sqrt{N_{\vec{q},\mu} + 
                         \frac12 \pm \frac12 } \right]_{\#}, \\
R_{e/a} & =  
      \nonumber \frac{M_{\vec{q}, \mu}}{\sqrt{\Omega}}
      \frac{1}{\sqrt{l_{cx} l_{vx}}}\times \\
      &\phantom{=} 
      \int_a^b \sqrt{\alpha_c(x) \alpha_v(x)}e^{-\mathfrak{f}_{e/a}(x)/ \hbar} dx 
\end{align}
\end{subequations}
with 
\begin{subequations}
\label{eq_elec_phon_int2}
\begin{align}
\begin{split}
\mathfrak{f}_{e/a} (x)  = & \phantom{+} \int_a^x \sqrt{2 m_{vx}(x')(E_{vx} - U(x'))} dx' \\
              & +  \int_x^b \sqrt{2 m_{cx}(x')(E_g + U(x') - E_{cx})} dx' \\
              & +  \iota Q_{e/a}x + \iota \hbar (k_{0vx}a - k_{0cx}b),
\end{split} \\
Q_{e/a}     = & \hbar(\pm q_x + k_{0cx} - k_{0vx} ), \\
\alpha_c(x) = & \frac{k_{cx} - k_{0cx}}{|\kappa_{cx}(x)|}
                 \frac{m_{cx}(x)}{m_{cx}},  \\
\alpha_v(x) = &  \frac{k_{vx} - k_{0vx}}{|\kappa_{vx}(x)|}
                 \frac{m_{vx}(x)}{m_{vx}} \text{ and } \\
N_{\vec{q}, \mu}  = & \frac{1}{
				\exp(\hbar \omega_{\vec{q}, \mu}) / k_BT) - 1}.
\end{align}
\end{subequations}
The overbar in eq.  (\ref{eq_elec_phon_int1}) indicates the use of the
expectation value $N_{\vec{q},\mu}$  for the phonon occupation number.
The  $\#$ in  eq.  (\ref{eq_elec_phon_int1})  specifies  the condition
$q_{y}  = \pm(k_{vy}  - k_{cy})$,  $q_z= \pm(k_{vz}  -  k_{cz})$. This
condition  is   obtained  from  the   fact  for  example   that  $\int
\exp(\iota(q_y -  (k_{cy} - k_{vy})))  dy = l_y \delta_{q_y,  k_{cy} -
k_{vy}}$  and  the  assumption   that  $M_{\vec{q},  \mu}$  is  weakly
dependent  on $\vec{q}$.   The upper  (lower)  sign in  $\pm$ in  eqs.
(\ref{eq_elec_phon_int1}),  (\ref{eq_elec_phon_int2})  corresponds  to
the first  (second) process in $e/a$  (i.e phonon emission/absorption)
in the transfer of an electron from the valence band to the conduction
band.    

The   integral   involving  $e^{-\mathfrak{f}_{e/a}(x)}$   is next
evaluated using the saddle point  method. Extending $x$ to the complex
plane $w$, we have
\begin{align}
\label{eq_saddlepoint}
\begin{split}
R_{e/a} = \frac{ M_{\vec{q}, \mu}}{\sqrt{\Omega}} 
         \sqrt{ \frac{2 \pi \hbar}{l_{cx}l_{vx}} } 
         \frac{\sqrt{\alpha_c(w_{\sigma}) \alpha_v(w_{\sigma}) }}{
         \sqrt{ \left|\frac{d^2 \mathfrak{f}_{e/a}(w_{\sigma})}{dw^2} \right|} }   
         { \exp \left(-\frac{\mathfrak{f}_{e/a}(w_{\sigma})}{\hbar}\right) }        
\end{split}
\end{align}
with    $\frac{d\mathfrak{f}_{e/a}(w_{\sigma})}{dw}   =   0$.     The   prefactor
$\sqrt{\alpha_{c}(x)\alpha_{v}(x)}$  in the  integral  is approximated
with its  value at the saddle  point. 

Next consider the processes $\{c \rightarrow v; e/a\}$, leading to
\begin{align}
\label{eq_elec_phon_int1_reverse}
\overline{\bra{v} \mathcal{W}_{e-ph} \ket{c}_{e/a}}
       & = 
      \sum_{q_x, \mu} \left[ R_{e/a}' \sqrt{N_{\vec{q},\mu} + 
                         \frac12 \pm \frac12 } \right]_{\#'}.
\end{align}
The $'$ denotes the reversal  of the transfer direction. $\#'$ implies
$q_{y} =  \mp(k_{vy} - k_{cy})$,  $q_z= \mp(k_{vz} -  k_{cz})$.  Using
the Hermiticity of $\mathcal{W}_{e-ph}$, we can show that $|R_{e/a}| =
|R_{a/e}'|$  for a  given pair  of  energies $E_v$,  $E_c$ and  phonon
energy $\omega_{\vec{q}, \mu}$. This relationship allows us to describe
all the four processes in terms of quantities derived for the two $\{v
\rightarrow c; e/a\}$ processes.

We now derive the net number of electrons, $N_t$, transferred per unit
time  from $v  \rightarrow c$  (including spin)  using  Fermi's golden
rule.  As mentioned earlier, each pair of intersecting complex valence
and  conduction bands  constitutes  a tunneling  path.   Based on  the
symmetry  of the  crystal,  there  can be  a  multiplicity of  $\nu_p$
different values  of $\vec{k}^{\parallel}$ within  the first Brillouin
zone (and  hence $\nu_p$ tunneling  paths) that give the  same complex
bands     $k^{\perp}(E;     \vec{k}^{\parallel})$    (see     Appendix
\ref{App_BZones}). Denoting $R_{e/a}$ evaluated in eq. (\ref{eq_saddlepoint})
along tunneling path $p$ as $R_{e/a, p}$, we have
\begin{align}
\label{eq_Nt}
\begin{split}
N_t = & 2 \sum_{p}  \sum_{\substack{ k_{vx}, k_{vy}, k_{vz} \\ 
                                     k_{cx}, k_{cy}, k_{cz}}}
       \sum_{q_x, \mu} \sum_{e/a}  \nu_p \frac{2 \pi}{\hbar} \Bigl|R_{e/a, p}\Bigr|^2_{\#}  
       \times \\
      & \Biggl[ \phantom{-}
                \Bigl(N_{\vec{q}, \mu} + \frac12 \pm \frac12 \Bigr)_{\#} \times
                f_v(E_v) \bigl( 1 - f_c(E_c) \bigr )  \\
      &   -     \Bigl(N_{\vec{q}, \mu} + \frac12 \mp \frac12 \Bigr)_{\#} \times
                f_c(E_c) \bigl( 1 - f_v(E_v) \bigr ) 
        \Biggr] \times \\
      & \Bigl[ \delta(E_c \pm \hbar \omega_{\vec{q}, \mu} - E_v ) \Bigr]_{\#}.  
\end{split}
\end{align}
where   $f_v(E_v),  f_c(E_c)$   are   the  Fermi
functions  evaluated  on the  two  sides.   As  described in  Appendix
\ref{App_Derivation}, the  current density $J =  -e N_t /  l_y l_z$ is
then
\begin{widetext}
\begin{align}
\label{eq_Jfinal}
\begin{split}
J = &  \sum_{p}
    \frac{e \nu_p}{2^{2.25} \pi^{2.5} \hbar^{8.5}}
      \sqrt{m_{cy}^p m_{cz}^p m_{vy}^p m_{vz}^p} \\
    & \times \sum_{\mu, e/a} 
    \frac{(m_{cx}^p (x_0^p) m_{vx}^p(x_0^p))^{1.25}}{
         (m_{cx}^p(x_0^p) +  m_{vx}^p(x_0^p))^{0.75}} 
    \frac{(E_g \pm \hbar \omega )^{0.25}}{
         \left|\kappa_{cx}(x_0^p) \kappa_{vx}(x_0^p)\right|} 
    \left[{M_{\vec{q}, \mu}}^2\right]_{*} 
    {\left[-\dfrac{dU}{dx}\right]_{x_0}^{-0.5}} \\
    & \times
    \int\limits_{E_{min} \pm \hbar \omega}^{E_{max}} dE_{vx} 
    T(E_{vx}) 
    \left[ \bar{E}_{\perp}^2 - 
     \bar{E}_{\perp} (\bar{E}_{\perp} + 
                         E_{vx} \mp \hbar \omega - E_{min})
          e^{-(E_{vx} \mp \hbar \omega - E_{min})/\bar{E}_{\perp}}
    \right] \times \\
    & \qquad \qquad \phantom{-} 
    \Biggl[ \Bigl(N_{\vec{q}, \mu} + \frac12 \pm \frac12 \Bigr)_{*} \times
                f_v(E_{vx}) 
                \bigl( 1 - f_c(E_{vx} \mp \hbar \omega) \bigr ) \\
    & \qquad \qquad   -     \Bigl(N_{\vec{q}, \mu} + \frac12 \mp \frac12 \Bigr)_{*} \times
                f_c(E_{vx} \mp \hbar \omega) 
                \bigl( 1 - f_v(E_{vx}) \bigr ) 
    \Biggr], 
\end{split}
\end{align}
\end{widetext}
\noindent with $T(E_{vx}) = e^{-\Lambda}$, 
\begin{subequations}
\begin{align}
\label{eq_Lambda}
\begin{split}
\Lambda  = & \frac{2}{\hbar} 
\Bigg[
     \int_a^{x_0^p} \sqrt{2 m_{vx}^p(x')(E_{vx} - U(x'))} dx'\\
     & + \int_{x_0^p}^{b_0} \sqrt{2 m_{cx}^p(x') 
     (E_g \pm  \hbar \omega -( E_{vx} - U(x') )} dx' 
\Bigg],
\end{split}
\end{align}
where the crossover point $x_0^p$ is a solution of  
\begin{equation}
\label{eq_x0}
\frac{ m_{vx}^p(x_0^p) }{ m_{cx}^p(x_0^p) } = 
\frac{(E_g \pm  \hbar \omega -( E_{vx} - U(x_0^p) ))}{(E_{vx} - U(x_0^p) )},
\end{equation}
and $U(b_0) = E_{vx} - (E_g \pm \hbar\omega)$. The scaling factor
\begin{equation}
\label{eq_EParallelBar}
\bar{E}_{\perp}  = {\hbar} \Bigg/
\int_{x_0^p}^{b_0}  \sqrt{\frac{2 m_{cx}(x')}{
E_g \pm \hbar\omega - ( E_{vx} - U(x') )   }} dx'    
\end{equation}
\end{subequations}
and the $*$ refers to the condition $q_x= \pm(k_{0vx}^p - k_{0cx}^p)$,
$q_y   =  \pm(k_{0vy}^p   -  k_{0cy}^p)$,   $q_z  =   \pm(k_{0vz}^p  -
k_{0cz}^p)$.      The      effective     masses     $m_{cx}^p(x_0^p)$,
$m_{vx}^p(x_0^p)$   and    the   wavevectors   $\kappa_{cx}^p(x_0^p)$,
$\kappa_{vx}^p(x_0^p)$ are obtained  from the complex bandstructure of
the   material.   Note   that   we   have  used   $\omega$   to   mean
$\omega_{\vec{q},\mu}|_{*}$ in order to avoid tedious notation.

\begin{figure}[!b]
 \centering
  \includegraphics[scale=1]{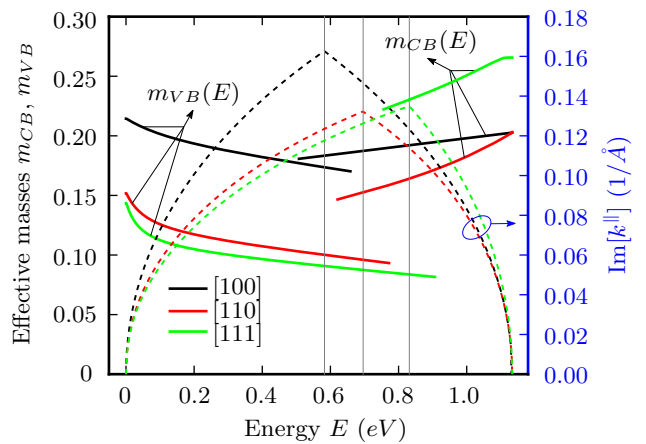}
\caption{Effective  masses $m_{CB}$,  $m_{VB}$ (solid  lines)  and imaginary
wavevectors      $\operatorname{Im}[k^{\parallel}]$     (dashed     lines)
corresponding  to the  tunneling path  that minimizes  area $\int_{Eg}
\operatorname{Im}[{k}^{\parallel}(E)] dE$ bounded by  the imaginary parts of the
valence and  conduction bands, in  silicon along the  $[100]$, $[110]$
and $[111]$ directions.}
\label{fig_EnergyDepMass}
\end{figure}

The expression  in eq. (\ref{eq_Jfinal}) is symmetric  with respect to
the  conduction  and  valence  band  masses.  The  term  $T(E_{vx})  =
e^{-\Lambda}$  is  independent  of  $E_{c\perp},  E_{v\perp}$  and  is
similar in  spirit to  the transmission coefficient  $T(E_x)$ computed
with a  transfer matrix method  in Ref.  \onlinecite{Pandey_TED_2010}.
To   correct   for  errors   introduced   by  neglecting   reflections
\cite{Huang_ChinJPhys_2008}     in    assuming    the     WKB    forms
eq. (\ref{eq_modified_VB}) and  eq. (\ref{eq_modified_CB}), we use the
transfer matrix method to  compute the term equivalent to $T(E_{vx})$.
We also  include the  velocity ratio ${(  k_{cx} -  k_{0cx}) m_{vx}}/{
(k_{vx} - k_{0vx})m_{cx}}$  in the formula for $T$.   We find that the
inclusion  of the  transfer matrix  method  changes the  current by  a
factor    approximately    between    $1    -   2$    (see    Appendix
\ref{App_TransferMatrix}). The deviation between a WKB calculation and
more accurate computational methods is known to be dependent on doping
(and hence  electric field).  A  similar trend of  WKB underestimating
the tunneling probability, as observed here, has been reported in Ref.
\onlinecite{Vandenberghe_JAP_2010} (in  the case  of BTBT in  a direct
bandgap   material   for  moderate   doping,   see   Fig.  6(c) therein)   and
Ref. \onlinecite{Mayer_JPCM_2010} (in the  case of tunneling through a
triangular barrier).

\begin{figure}[t]
 \centering
  \includegraphics[scale=0.9]{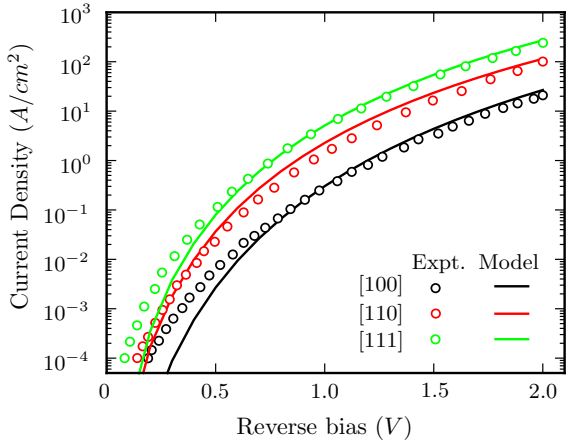}	
\caption{Comparison  between our model  and experimental  data (doping
$Q2$  of  Ref.  \onlinecite{Solomon_DRC_2009}).  Note  that  the  term
$T(E_{vx})$ in   eq. (\ref{eq_Jfinal}) has been  computed using transfer
matrices.}
\label{fig_ExptCompare}
\end{figure}

\section{Results}
\label{Sec_Results}
We    now     test    our    model     against    experimental    data
\cite{Solomon_DRC_2009} available  for BTBT  in silicon. This  data is
unique in  that the same  doping profile has  been used to  study BTBT
along the  $[100]$, $[110]$ and  $[111]$ directions. We  implement our
model in the open  source drift-diffusion based TCAD code \emph{pyEDA}
\cite{pyEDA}.   We also modify  the \emph{pyEDA}  code to  include the
effects  of degenerate  doping and  incomplete ionization  of dopants.
Information regarding  the phonon energies,  modes and electron-phonon
deformation       potentials      are       taken       from      Ref.
\onlinecite{Fischetti_TED_2007}.  Owing  to the conservation condition
$*$ in eq.  (\ref{eq_Jfinal}), we  only require the phonon modes along
the $\Delta \equiv \Gamma-X$ direction in silicon. To summarize, there
is  one longitudinal  optical (LO)  and  acoustic (LA)  mode, and  two
doubly  degenerate transverse  optical (TO)  and acoustic  (TA) phonon
modes  with   energies  $61.2$,  $47.4$,  $59.0$,   $19.0$  $meV$  and
deformation  potentials $6.0$,  $5.9$,  $6.0$, $5.9$  $\times 10^8  \;
eV/cm$ respectively. The  density $\rho_s = 2.328 \;  g/cm^3$ is taken
from  Ref.   \onlinecite{Tanaka_SSE_1995}. The  TA  mode provides  the
largest contribution to the  BTBT current (for example, $\approx 83\%$
of the  total current at  reverse bias of  $2.0 \; V$ for  the $[110]$
direction  in  the  device  considered  here.) In  order  to  simplify
computation, we  restrict ourselves to the $\nu$  tunneling paths (and
hence values  of $\vec{k}^{\perp}$) that minimize  the area $\int_{Eg}
\operatorname{Im}[{k}^{\parallel}(E)]  dE$  bounded  by the  imaginary
parts  of the  valence  and conduction  bands  involved in  tunneling.
Other tunneling paths enclose much larger areas and are hence expected
to contribute negligibly  to tunneling current, due to  the term $e^{-
\Lambda}$ in  eq.  (\ref{eq_Jfinal}).  Further, these  $\nu$ paths all
happen  to originate  from  the  valence band  for  light holes.   The
multiplicity $\nu = 4, 2,  6$ for transport along the $[100]$, $[110]$
and  $[111]$ directions  respectively  (Appendix \ref{App_BZones})  in
silicon.   Fig.  \ref{fig_EnergyDepMass}  shows  the energy  dependent
effective  mass computed  using an  $sp^3d^5s^*$ tight  binding scheme
\cite{Ajoy_DRC_2011}       and        parameters       from       Ref.
\onlinecite{Boykin_PRB_2004}.        The       invariant       product
$m_{cx}m_{cy}m_{cz}$ is $0.891 m_0 \times 0.201 m_0 \times 0.201 m_0$,
written using  a coordinate  system aligned with  the major  and minor
axes of any one of  the six conduction band ellipsoids. Similarly, the
product  $m_{vx}m_{vy}m_{vz}$ is  $0.214 m_0  \times 0.152  m_0 \times
0.144 m_0$, corresponding  to the effective masses of  the light holes
$m_{lh,[100]}$,  $m_{lh, [110]}$  and $m_{lh,[111]}$  along  the three
orthogonal $[100]$, $[110]$  and $[111]$ directions respectively.  The
value  of $m_{cy}m_{cz}$ and  $m_{vy}m_{vz}$ in  eq. (\ref{eq_Jfinal})
are obtained from these invariant products and the values of $m_{cx}$,
$m_{vx}$  in Fig.   \ref{fig_EnergyDepMass} at  the band  edges.  Fig.
\ref{fig_ExptCompare} shows  that the results of our  model agree very
well with  the experimental  data.  We have  assumed a bandgap  $E_g =
0.92 \; eV$, corresponding to a bandgap narrowing of $\sim 0.2 eV$, by
fitting the results of our  model with the experimental data (We found
this to  be a better  strategy than calculating the  bandgap narrowing
apriori, since the value of  bandgap narrowing is dependent on doping,
which is  non-uniform for  the devices we  have considered  here.  The
model for BTBT that we  have derived assumes a uniform bandgap.)  This
value of narrowing is consistent  with studies on bandgap narrowing in
space  charge regions  \cite{Chen_SSE_1989,  Lowney_SSE_1985} for  the
doping levels  considered here.  Further,  based on a  result obtained
using  $\vec{k}\cdot\vec{p}$  theory  \cite{Miller_Book_2008} that  $m
\propto E_g$,  we have scaled all  the effective masses  by the factor
$E_g / E_{g0}$, where $E_{g0}$  is the bandgap for moderate doping. At
low   values   of  reverse   bias,   our   model  underestimates   the
experimentally  observed value  of  current (for  e.g., for  transport
along   $[100]$,   at   $V_{bias}   =   0.25V$,   $I_{expt.}   \approx
3\times10^{-4}  \;  A/cm^{2}$ whereas  $I_{model}  \approx 4.7  \times
10^{-5}  \; A/cm^{2}$).  It is  likely  that some  other mechanism  of
current transport  (such as tunneling via traps,  or SRH recombination
via  traps)   could  possibly  explain  the   difference  between  our
simulations and experimental data at small values of reverse bias (see
for e.g. Fig. 7  of Ref. \onlinecite{Schenk_SSE_1993}). It is possible
that the trap distribution/energies  are different in the experimental
samples that  we have compared  our model against for  transport along
the  different  directions, leading  to  a  better  match between  the
experimental data and the model for the $[110]$ direction. A detailed
analysis of this deviation could be the focus of future work.

\section{Error due to parabolic approximation of complex bands}
\label{Sec_Parabolic}
\begin{figure*}[!t]
 \centering
  \includegraphics[scale=1]{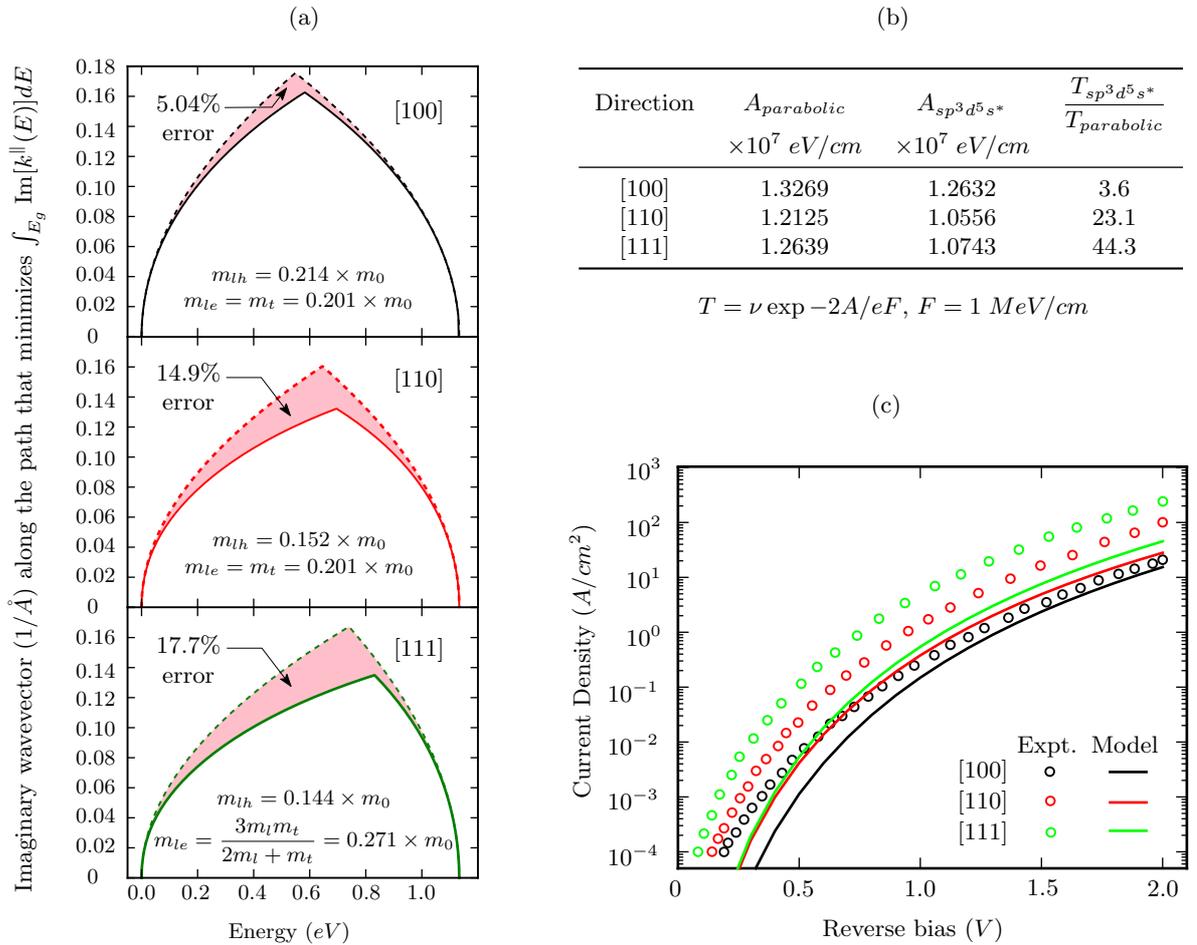}
\caption{(a) Parabolic approximation (dashed) to complex bands (solid)
showing    the    errors     in    estimating    $A    =    \int_{E_g}
\operatorname{Im}[k^{\parallel} (E)]dE$. (b) Comparison of results for the
case of a uniform field F using a simple WKB expression (c) Comparison
of results using our BTBT model for parabolic complex bands. Note that
we use the TM method with  the same bandgap (and scaling of masses) as
in Fig.  3.
}
\label{fig_ParabolicApproximation}
\end{figure*}
We now  demonstrate the inadequacy of using  a parabolic approximation
to  the  complex bandstructure  while  computing  BTBT currents.  Fig.
\ref{fig_ParabolicApproximation}(a)  shows  a parabolic  approximation
(dashed lines) to  the complex bands (solid lines,  obtained from an
$sp^3d^5s^*$ calculation) along the tunneling path that minimizes $A =
\int_{E_g} \operatorname{Im}[k^{\parallel}(E)]dE$.   The curvatures of
the imaginary and  real bands are identical at  the band extrema.  The
values    of   $m_l$,    $m_t$    and   $m_{lh}$    are   from    Ref.
\onlinecite{Boykin_PRB_2004}.  The  expressions for $m_{le}$  are from
Ref.    \onlinecite{Pandey_TED_2010},   based   on   the   theory   in
Ref. \onlinecite{Rahman_JAP_2005}.  The  areas $A$ in the $sp^3d^5s^*$
method   and    parabolic   approximations   are    listed   in   Fig.
\ref{fig_ParabolicApproximation}(b);  the errors  due  to a  parabolic
approximation with  respect to the $sp^3d^5s^*$  results are indicated
in Fig.  \ref{fig_ParabolicApproximation}(a).   Note that the error is
largest  along  the  $[111]$  direction.   A  simple  result  for  the
transmission  (setting the phonon  energy to  $0$, assuming  a uniform
field $F$, and using  a WKB approximation) gives \cite{Laux_IWCE_2009}
$T =  \nu e^{-2A/eF}$,  where $\nu$ is  the multiplicity  of tunneling
paths.  As indicated  in Fig. \ref{fig_ParabolicApproximation}(b), the
parabolic  approximation  underestimates the  tunneling  current by  a
factor  of   $44.3$  along  the  $[111]$   direction.   Finally,  Fig.
\ref{fig_ParabolicApproximation}(c)  shows the  results  of using  the
parabolic approximation  in our BTBT  model (as usual,  $T(E_{vx})$ is
computed  using transfer  matrices).   We use  the  same bandgap  (and
scaling  of masses)  as in  Fig. \ref{fig_ExptCompare}.   Clearly, the
parabolic approximation  does not capture the  measured data. Further,
it significantly underestimates the difference between the currents in
the $[111]$  and $[100]$  directions.  We would  like to  clarify that
though the choice of effective energy gap  can  increase or  decrease the
absolute values of the current levels, it cannot correctly predict the
difference  in currents  between the  $[111]$ and  $[100]$ directions.
This  can been  seen  from Fig.   \ref{fig_ParabolicApproximation}(a),
where error in the action for tunneling along the $[111]$ direction is
significantly greater than that along the $[100]$ direction.

\section{Conclusion}
\label{Sec_Conclusion}
In  conclusion,  we  have  presented  a multiscale  model  for  phonon
assisted BTBT  that accounts for the complex  bandstructure within the
bandgap  of  an  indirect  semiconductor.   We  have  shown  that  the
predictions of this model compare very well with experimental data for
BTBT  in silicon  along different  orientations.  We  have  shown that
including the  effect of non-parabolic  complex bands is  important to
capture  the correct  difference between  tunneling  currents observed
along  the  $[100]$, $[110]$  and  $[111]$  directions. The  framework
presented   here   can   be    used   to   modify   Tanaka's   results
\cite{Tanaka_SSE_1994} on BTBT across  a direct bandgap to include the
effect of an energy dependent  effective mass.  Such an extension will
find  application in  treating BTBT  in materials  such  as germanium,
where the  direct bandgap  is only about  $0.15 \;eV$ larger  than the
indirect bandgap.

\appendix
\section{Description of wavefunctions and energies}
\label {App_Tanaka}
\begin{table*}[!t]
\caption{Description   of   wavefunctions   assumed,  following   Ref.
\onlinecite{Tanaka_SSE_1994}.}  
\label{tab_summary}
  \begin{center}
  {  
  \begin{ruledtabular}  
  \begin{tabular}{ccll}  
       & Region  
       & \multicolumn{1}{c}{Wavefunction} 
       & \multicolumn{1}{c}{Energy}   \\ 
   \cline{2-4}   
   & & $\psi_v(x,y,z)  =  u_{vx}(x) u_{vy}(y) u_{vz}(z)$ & $E_v = E_{vx} - E_{v \perp}$  \\ \\
   \multirow{3}{1cm}[-0.5cm]{VB}  &  
      $-\infty < x < b$ & 
      $  u_{vy}(y) =  e^{\iota k_{vy} y} / \sqrt{l_{y}}, \;
         u_{vz}(z) =  e^{\iota k_{vz} z} / \sqrt{l_{z}}
      $ &
      $ E_{v \perp} =
        \hbar^2  (k_{vy} - k_{0vy})^2 / 2 m_{vy} 
        + \hbar^2  (k_{vz} - k_{0vz})^2 / 2 m_{vz}
      $ \\ \\
      &
      $-\infty < x < a$ &
      $ u_{vx}(x) = e^{\iota k_{vx} (x - a)} / \sqrt{l_{vx}}  $ &
      $ E_{vx} = E_{max} 
      - \hbar^2 (k_{vx} - k_{0vx})^2/ 2 m_{vx} $ \\ \\
      
      &
      $ a  < x < b$ &
      $ u_{vx}(x)$ --- eq. (\ref{eq_modified_VB})  &
      $ \begin{aligned}
        \left[-\frac{\hbar^2}{2m_{vx}} 
              \left(-\iota \frac{d}{dx} - k_{0vx} \right)^2 +
              U(x)\right]u_{vx}(x) 
        = E_{vx}u_{vx}(x)
        \end{aligned}
       $           \\ \\
         
   \hline  
       
   & & $\psi_c(x,y,z)  =  u_{cx}(x) u_{cy}(y) u_{cz}(z) $ &$E_c = E_{cx} + E_{c \perp}$ \\ \\
   \multirow{3}{1cm}[-1cm]{CB}  & 
      $a < x < \infty$ & 
      $  u_{cy}(y) =  e^{\iota k_{cy} y} / \sqrt{l_{y}}, \;
         u_{cz}(z) =  e^{\iota k_{cz} z} / \sqrt{l_{z}}
      $ &
      $ E_{c \perp} = 
        \hbar^2  (k_{cy} - k_{0cy})^2 / 2 m_{cy} 
        + \hbar^2(k_{cz} - k_{0cz})^2 / 2 m_{cz}
      $ \\  
      &   
      $ a  < x < b$ &
      $ u_{cx}(x)$ --- eq. (\ref{eq_modified_CB})  &
      $ \begin{aligned}
        \left[ \frac{\hbar^2}{2m_{cx}} 
              \left(-\iota \frac{d}{dx} - k_{0cx} \right)^2 +
              E_g + U(x)\right]u_{cx}(x) 
          =  E_{cx}u_{cx}(x)
        \end{aligned}  
      $  \\ \\
      &
      $ b  < x < \infty$ &
      $ u_{cx}(x) = e^{\iota k_{cx} (x - b)} / \sqrt{l_{cx}}  $ &
      $ E_{cx} = E_{min} + \hbar^2 (k_{cx} - k_{0cx})^2/ 2 m_{cx} $ \\ \\

  \end{tabular}
  \end{ruledtabular} 
  }
  \end{center}
\end{table*}
The wavefunctions within the  effective mass approximation are written
as  $\Psi_v(x,y,z)   =  u_{vx}(x)\allowbreak  u_{vy}(y)u_{vz}(z)$  and
$\Psi_c(x,y,z) =  u_{cx}(x)u_{cy}(y)u_{cz}(z)$ for an  electron in the
valence and  conduction bands respectively. The extents  of the device
in   the   $y,z$   directions    are   $l_y,   l_z$.   As   shown   in
Fig.  \ref{fig_Tanaka}, the  component of  the energy  of  an electron
corresponding  to   its  motion  in  the  $yz$   plane  is  designated
$E_{v \perp}$  in the  valence band and  $E_{c \perp}$  in the
conduction band.  Note that $E_{v \perp}, E_{c \perp} \ge 0.$

Further,   following   Ref.  \onlinecite{Tanaka_SSE_1994},   $u_{vy}$,
$u_{cy}$ and $u_{vz}$,  $u_{cz}$ are plane waves.  On  the other hand,
$u_{vx}$  and $u_{cx}$  are  assumed  to be  plane  waves outside  the
classical turning points (i.e.  $x<a$ and $x>b$).  This corresponds to
making      the     approximation      (see      Fig.     1(b)      of
Ref.  \onlinecite{Tanaka_SSE_1994})  that the  energy  bands are  flat
until $x =  a^-$ (with the valence band edge  at $E_{max}$) and beyond
$x  =  b^+$ (with  the  conduction  band  edge at  $E_{min}$).   Table
\ref{tab_summary} summarizes the expressions for the wavefunctions and
energies in different regions.
\begin{figure*}[!ht]
 \centering
	\includegraphics[scale=1]{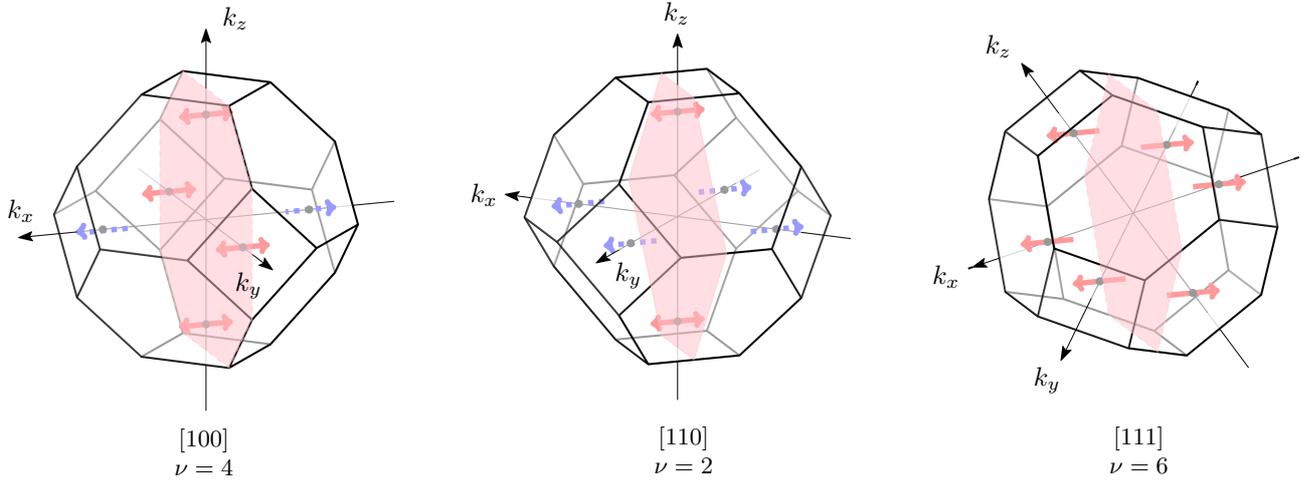}
\caption{Multiplicity  $\nu$   of  tunneling  paths   in  silicon  along 
the $[100]$, $[110]$ and $[111]$ directions. The plane
$\vec{k}^{\parallel} = \vec{0}$ is shown shaded. The red (solid) paths provide a higher
tunneling probability than the blue (dashed) paths.}
\label{fig_BrillouinZoneFigs}
\end{figure*}
\section{Derivation of tunneling current}
\label{App_Derivation}
\setcounter{equation}{0}
The  summations in eq. (\ref{eq_Nt})  are  first  converted into
integrals, for  e.g.  $\sum_{k_{cx}} \rightarrow  \frac{l_{cx}}{2 \pi}
\int  dk_{cx}$, $\sum_{k_{cy}}  \rightarrow  \frac{l_{y}}{2 \pi}  \int
dk_{cy}$. Further,  the integrals are  rewritten in terms  of energies
using  the relationship  between $k$  and  $E$ outside  the region  of
tunneling, for e.g.  $\int dk_{cx} = \frac{m_{cx}}{\sqrt{2}\hbar} \int
\frac{dE_{cx}}{\sqrt{E_{cx}}}$,  $\int dk_{cy}  dk_{cz} =  \frac{2 \pi
\sqrt{m_{cy}m_{cz}}}{\hbar^2}  \int  dE_{c  \perp}$. In  order  to
determine  the limits of  integration, we  impose the  conditions that
$E_c \ge  E_{min}$ and $E_v \le  E_{max}$ .  By  definition, $E_c, E_v
\ge  0$. Anticipating  the  physical reality  that  tunneling will  be
dominated   by   states  with   $E_{\perp}   =  E_{c\perp}   +
E_{v\perp}  \to 0$,  we also  impose conditions  that  $E_{cx} \ge
E_{min}$  and  $E_{vx}  \le  E_{max}$  .  This  gives  the  limits  of
integration.  The current density $J = -eN_t /l_yl_z$ is
\begin{widetext}
\begin{align}
\label{eq_J1}
\begin{split}
J  = & \sum_{p} 
\frac{e \nu_p}{ 4 \pi^3 \hbar^8} 
\sqrt{m_{cy}^p m_{cz}^p m_{vy}^p m_{vz}^p} 
\sum_{\mu, e/a} 
\frac{m_{cx}^p(w_{\sigma}^p) m_{vx}^p(w_{\sigma}^p)}{
|\kappa_{cx}^p(w_{\sigma}^p)\kappa_{vx}^p(w_{\sigma}^p)|} 
  \int dq_x
  \int\limits_{E_{min} \pm \hbar \omega}^{E_{max}} dE_{vx}
  \int\limits_{0}^{E_{vx} \mp \hbar \omega -E_{min}} dE_{v\perp} 
  \int\limits_{0}^{E_{max} \mp \hbar \omega - E_{min}} dE_{c\perp} \\
& \times
  \int\limits_{E_{min}}^{E_{max} \mp \hbar \omega - E_{c\perp} }dE_{cx} 
  \left[ M_{\vec{q}, \mu}^2 \right]_{\#} 
  \left| {\frac{d^2 \mathfrak{f}_{e/a}(w_{\sigma}^p)}{dw^2}}\right|^{-1} 
  \left| \; \exp \left (-\frac{2 \mathfrak{f}_{e/a}(w_{\sigma}^p)}{\hbar} \right) \right| 
  \delta \left(E_{cx} - (E_{vx} \mp \hbar \omega - E_{c\perp} - E_{v\perp}) \right)\\
& \times
\Biggl[ \Bigl(N_{\vec{q}, \mu} + \frac12 \pm \frac12 \Bigr)_{\#} 
                f_v(E_{vx} - E_{v\perp}) 
                \bigl( 1 - f_c(E_{cx} + E_{c\perp}) \bigr )
              - \Bigl(N_{\vec{q}, \mu} + \frac12 \mp \frac12 \Bigr)_{\#}
                f_c(E_{cx} + E_{c\perp}) 
                \bigl( 1 - f_v(E_{vx} - E_{v\perp}) \bigr ) \Biggr] 
\end{split}
\end{align}
\end{widetext}
where the summation over $q_x$ has also been converted to an integral.
$E_{cx}$   is  eliminated   from eq. (\ref{eq_J1})   due  to   the  delta
function. Further  simplification requires making  the assumption that
only states  with small values of  $E_{c\perp}$, $E_{v \perp}$
and hence $E_{\perp} = E_{c\perp} + E_{v\perp}$ contribute
significantly    to    tunneling.     Tanaka    \cite{Tanaka_SSE_1994}
approximates an integral of the form $\int_{0}^{X_0} \int_{0}^{Y_0}F(X
+Y) dX  dY \approx \int_{0}^{Y_0}V  F(V) dV$ where  $V = X+Y$  (and $X
\equiv E_{c\perp}$,  $Y \equiv  E_{v\perp}$) for the  case that
$F(V)$  is   significant  only   for  small  $V$.   However,  Tanaka's
expressions do not include the Fermi functions as in eq. (\ref{eq_J1}). We
drop $E_{v\perp}$ in  the arguments  of the  Fermi functions
$f_c, f_v$ and write
\begin{align}
\begin{split}
\label{eq_J2}
J  = & \sum_{p} 
\frac{e \nu_p}{ 4 \pi^3 \hbar^8} 
\sqrt{m_{cy}^p m_{cz}^p m_{vy}^p m_{vz}^p} 
\sum_{\mu, e/a} 
\frac{m_{cx}^p(w_{\sigma}^p) m_{vx}^p(w_{\sigma}^p)}{
|\kappa_{cx}^p(w_{\sigma}^p)\kappa_{vx}^p(w_{\sigma}^p)|} \\
& \times \int dq_x
  \int\limits_{E_{min} \pm \hbar \omega}^{E_{max}} dE_{vx}
  \int\limits_{0}^{E_{vx} \mp \hbar \omega -E{min}} E_{\perp} dE_{\perp}\\
& \times 
  \left[{M_{\vec{q}, \mu}}^2 \right]_{\#}
  \left| {\frac{d^2 \mathfrak{f}_{e/a}(w_{\sigma}^p)}{dw^2}}\right|^{-1} 
  \left| \; \exp \left(-\frac{2 \mathfrak{f}_{e/a}(w_{\sigma}^p)}{\hbar} \right) \right| \\
& \times
  \Biggl[    \Bigl(N_{\vec{q}, \mu} + \frac12 \pm \frac12 \Bigr)_{\#} \times
                f_v(E_{vx}) 
                \bigl( 1 - f_c(E_{vx} \mp \hbar \omega) \bigr ) \\
& \quad -     \Bigl(N_{\vec{q}, \mu} + \frac12 \mp \frac12 \Bigr)_{\#} \times
                f_c(E_{vx} \mp \hbar \omega) 
                \bigl( 1 - f_v(E_{vx}) \bigr ) 
    \Biggr].
\end{split}
\end{align}
Finally,  following Ref.  \onlinecite{Tanaka_SSE_1994}, $w_{\sigma}^p$
and   hence  $\mathfrak{f}_{e/a}(w_{\sigma}^p)$   are   functions  of   $Q_{e/a},
E_{\perp}$. We expect that  the dominant contribution to tunneling
will  be for  $Q_{e/a}  = 0$,  $E_{\perp}  = 0$.   We thus  expand
$\mathfrak{f}_{e/a}(w_{\sigma}^p)$ using a Taylor approximation
\begin{equation}
\label{eq_Taylor}
\begin{split}
\mathfrak{f}_{e/a}(w_{\sigma}^p) \equiv & \mathfrak{f}_{e/a}(Q_{e/a}, E_{\perp}) = 
 \mathfrak{f}_{e/a}(0,0) + 
\frac{\partial  \mathfrak{f}_{e/a}(0,0)}{ \partial Q_{e/a}} Q_{e/a} \\ 
& + \frac{1}{2}
\frac{\partial^2 \mathfrak{f}_{e/a}(0,0)}{ \partial Q_{e/a}^2}Q^2_{e/a} + 
\frac{\partial  \mathfrak{f}_{e/a}(0,0)}{ \partial E_{\perp}} E_{\perp}.
\end{split}
\end{equation}
To  determine the  coefficients in  the  above equation,  we make  the
approximation  that   $m_{cx}^p(x)$,  $m_{vx}^p(x)$  are   gently  varying
functions  of $x$, and  hence ignore  their spatial  derivatives. This
allows    reuse   of    many   of    the   expressions    derived   in
Ref. \onlinecite{Tanaka_SSE_1994} with minor modifications. Then $x_0^p =
w_{\sigma}^p( 0, 0)$ is a solution of the equation
\begin{subequations}
\begin{equation}
\begin{split}
\label{eq_xOapp}
\sqrt{2 m_{vx}^p(x) (E_{vx} - U(x_0^p) )} = \\
\sqrt{2 m_{cx}^p(x) (E_g \pm  \hbar \omega -( E_{vx} - U(x_0^p) )}
\end{split}
\end{equation}
and represents the point of intersection of the imaginary parts of the
complex valence and conduction bands. We have 
\begin{align}
\Lambda  = &  \frac{2 \mathfrak{f}_{e/a}(0,0)}{\hbar} 
\end{align} 
which gives eq. (\ref{eq_Lambda}). The coefficient 
$\frac{\partial  \mathfrak{f}_{e/a}(0,0)}{ \partial  Q_{e/a}} = \iota x_0^p$ is 
purely imaginary and hence can be ignored. Further, 
\begin{equation}
\label{eq_coeff_d2fdQ2}
\frac{\partial^2 \mathfrak{f}_{e/a}(0,0)}{ \partial Q_{e/a}^2} = 
2 \frac{\sqrt{ 2 m_{rx_0}(E_g \pm \hbar \omega)}}
{m_{cx}^p(x_0^p) +  m_{vx}^p(x_0^p)} \Bigg/ {\left[-\dfrac{dU}{dx}\right]_{x_0^p}}
\end{equation} 
where $ m_{rx_0}$ is a reduced mass given by
\begin{equation}
\label{eq_reducedmass}
 m_{rx_0} = \frac{m_{cx}^p(x_0^p) m_{vx}^p(x_0^p)}{m_{cx}^p(x_0^p) + m_{vx}^p(x_0^p)}.
\end{equation}
Also, 
\begin{equation}
\label{eq_df_dE}
\frac{\partial  \mathfrak{f}_{e/a}(0,0)}{ \partial E_{\perp}} = 
\frac{1}{2}\int_{x_0^p}^{b_0} 
\sqrt{\frac{2 m_{cx}^p(x')}{
E_g \pm \hbar\omega - ( E_{vx} - U(x') )   }}
\end{equation}
and
\begin{equation}
\label{eq_d2f_dw2}
{\frac{d^2 \mathfrak{f}_{e/a}(0,0)}{dw^2}} = 
\frac{m_{cx}^p(x_0) + m_{vx}^p(x_0)}{
\sqrt{ 2 m_{rx_0}(E_g \pm \hbar \omega)}}
{\left[-\dfrac{dU}{dx}\right]_{x_0^p}}.
 \end{equation}
\end{subequations}
Finally, from  eqs. (\ref{eq_J2}), (\ref{eq_Taylor}), (\ref{eq_xOapp})
-  (\ref{eq_d2f_dw2}), we  get eq.  (\ref{eq_Jfinal}).  Note  that the
Gaussian  integral over  $q_x$ converts  the condition  $\#$  into the
condition $*$  since $\int [M_{\vec{q}, \mu}]_{\#} \exp  (- \beta Q^2)
dq_x \approx  [M_{\vec{q}, \mu}]_{*} \sqrt{\pi/ \beta}$  by the method
of steepest descent.

\section{Multiplicity of tunneling paths}
\label{App_BZones}
\setcounter{equation}{0}
The conduction  band minima  in silicon are  along the  six equivalent
$\langle     100    \rangle$     directions.    As     described    in
Ref. \onlinecite{Ajoy_DRC_2011}, the positions  of these valleys are used to
determine the  values of $\vec{k}^{\perp}$ to  compute the complex
bands $k^{\parallel}  (E; \vec{k}^{\perp})$. The  value $k^{\parallel} (E;
\vec{k}^{\perp})$ represents the magnitude of the component of the
wavevector,  parallel  or  antiparallel  to the  transport  direction,
oriented  along   the  direction  of  an  arrow   through  the  valley
corresponding     to     $\vec{k}^{\perp}$,     as    shown     in
Fig.  \ref{fig_BrillouinZoneFigs}.  Paths  having the  same $k^{\parallel}
(E; \vec{k}^{\perp})$  are shown in the same  color.  Further, the
paths  shown solid in red  have complex  bands that  enclose a  smaller area
$\int_{E_g}   \operatorname{Im}[k^{\parallel}(E)]dE$    bounded   by   the
imaginary parts of  the valence and conduction bands  than those shown
dashed in blue.  It  is necessary to consider valleys that  lie on both sides
of   the   $\vec{k}^{\parallel}   =   \vec{0}$   plane   (shown   shaded).
Fig. \ref{fig_BrillouinZoneFigs} gives  the multiplicity $\nu$ for the
$[100]$, $[110]$ and $[111]$ directions. 

\section{Transfer Matrix Method as an improvement over the WKB approximation}
\label{App_TransferMatrix}
\setcounter{equation}{0}
\begin{figure}[!t]
 \centering
  \includegraphics[scale=1]{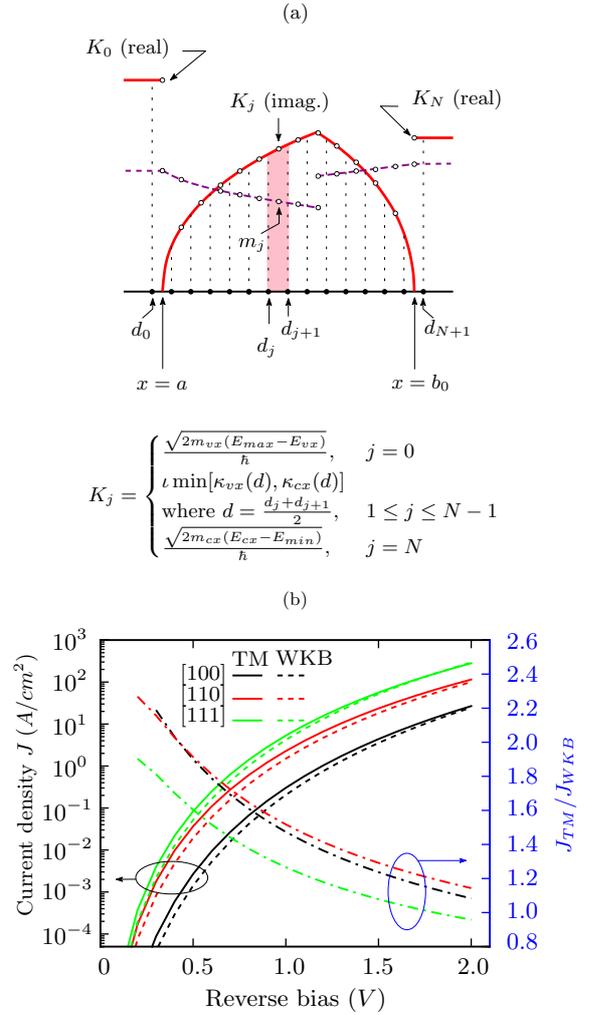}	
\caption{(a)  Description  of the  transfer  matrix  (TM) method.  (b)
Comparison of using  the transfer matrix and WKB  methods in the final
expression  for BTBT  current for  the devices  described in  eq. (\ref{eq_Jfinal}).}
\label{fig_TransferMatrix}
\end{figure}
In order to  use the transfer matrix method, we  define points $d_j$ ,
$j    =    0,    1,    \dots    ,    N    +    1$    as    shown    in
Fig. \ref{fig_TransferMatrix}(a), so that the values of the wavevector
are available at the midpoints of intervals $[d_j , d_{j+1} ]$, $0 \le
j \le  N$ . There are  $N$ interfaces $(1 \le  j \le N  )$ between the
classical turning  points $x = a$  and $x =  b_0$ . For each  of these
interfaces, we have  \cite{Pandey_TED_2010} a transfer matrix $[M_j]$,
given by
\begin{align}
[M_j] = \frac{1}{2 K_j m_{j+1}} 
        \begin{bmatrix}
          C_j e^{  \iota (K_{j+1} - K_{j}) d_j} & 
          D_j e^{ -\iota (K_{j+1} + K_{j}) d_j} \\ 
          D_j e^{  \iota (K_{j+1} + K_{j}) d_j} & 
          C_j e^{ -\iota (K_{j+1} - K_{j}) d_j} 
        \end{bmatrix}  
\end{align}
with  $C_j =  K_j m_{j+1}  + K_{j+1}  m_j$ and  $D_j =  K_j  m_{j+1} -
K_{j+1} m_j$. $K_j$ is defined  in Fig. \ref{fig_TransferMatrix}(a);
$m_j$ is similarly evaluated from $m_{vx}(x)$ or $m_{cx}(x)$, based on
whether the wavevector corresponds  to the valence or conduction bands
respectively. Note  that $E_{cx}  = E_{vx} \mp  \hbar \omega$  for the
situation   $Q_{e/a}  =   0$,   $E_{\perp}  =   0$  described   in
eq. (\ref{eq_Taylor}). The transmission $T(E_{vx})$ is then
\begin{subequations}
\begin{align}
T(E_{vx}) & =  \frac{1}{\big| [M]_{1,1} \big|^2} \times
              \frac{k_{cx} - k_{0cx}}{k_{vx} - k_{0vx}}
              \frac{m_{vx}}{m_{cx}}, \text{ where} \\
   [M]   & = [M_1][M_2] \dots [M_3]   .
\end{align}
\end{subequations}
Also note that $K_0 \equiv k_{vx}  - k_{0vx}$ and $K_N \equiv k_{cx} -
k_{0cx}$  based on  the  assumption  that the  energy  bands are  flat
outside the classical turning points.

A  comparison   of  using  the  transfer  matrix   method  to  compute
$T(E_{vx})$  instead of  using  the WKB  result  $T (E_{vx}  ) =  e^{-
\Lambda}$    in    eq.    (\ref{eq_Jfinal})    is   shown    in    Fig.
\ref{fig_TransferMatrix}(b).  The transfer  matrix  method predicts  a
higher current over most of the bias range.

\section*{Acknowledgment}
A. Ajoy wishes  to thank IBM India for  financial support. The authors
wish to thank Dr. Rajan Pandey (IBM Bangalore) and G. Vijayakumar (IIT
Madras) for useful discussions.


%

\end{document}